\catcode`\@=11
\newif\if@fewtab\@fewtabtrue

{\count255=\time\divide\count255 by 60
\xdef\hourmin{\number\count255}
\multiply\count255 by-60\advance\count255 by\time
\xdef\hourmin{\hourmin:\ifnum\count255<10 0\fi\the\count255}}
\def\ps@draft{\let\@mkboth\@gobbletwo
    \def\@oddhead{}
    \def\@oddfoot
       {\hbox to 7 cm{$\scriptstyle Draft\ version:\ \draftdate$
       \hfil}\hskip -7cm\hfil\rm\thepage \hfil}
    \def\@evenhead{}\let\@evenfoot\@oddfoot}

\def\ceqno{\global\@fewtabfalse
    \ifcase\@eqcnt \def\@tempa{& & &}\or \def\@tempa{& &}
      \or \def\@tempa{&}
      \or\def\@tempa{}\fi\@tempa
{\rm(\theequation)}}

\def\aeqno#1{\global\@fewtabfalse
    \ifcase\@eqcnt \def\@tempa{& & &}\or \def\@tempa{& &}
      \or \def\@tempa{&}
      \or\def\@tempa{}\fi\@tempa
{\rm(\theequation,#1)}}

\def\label#1{\ifnum\draftcontrol=1
 \global\def\draftnote{$\scriptstyle #1$}\fi
 \@bsphack\if@filesw {\let\thepage\relax
   \def\protect{\noexpand\noexpand\noexpand}%
\xdef\@gtempa{\write\@auxout{\string
      \newlabel{#1}{{\@currentlabel}{\thepage}}}}}\@gtempa
   \if@nobreak \ifvmode\nobreak\fi\fi\fi
  \@esphack}

\def\alabel#1#2{\label{#1}\global\@fewtabfalse
    \ifcase\@eqcnt \def\@tempa{& & &}\or \def\@tempa{& &}
      \or \def\@tempa{&}
      \or\def\@tempa{}\fi\@tempa
{\hbox to 3cm{\phantom{\rm(\theequation,#2)}
\draftnote \hfil}\hskip -3cm {\rm(\theequation,#2)}}}

\def\clabel#1{\label{#1}\global\@fewtabfalse
    \ifcase\@eqcnt \def\@tempa{& & &}\or \def\@tempa{& &}
      \or \def\@tempa{&}
      \or\def\@tempa{}\fi\@tempa
{\hbox to 3cm{\phantom{\rm(\theequation)}
\draftnote \hfil}\hskip -3cm{\rm(\theequation)}}}

\def\eqnarray{\def\draftnote{{}}\global\@fewtabtrue
\stepcounter{equation}\let\@currentlabel=\theequation
\global\@eqnswtrue
\global\@eqcnt\z@\tabskip\@centering\let\\=\@eqncr
$$\halign to \displaywidth\bgroup\@eqnsel\hskip\@centering\@eqcnt\z@
  $\displaystyle\tabskip\z@{##}$&\global\@eqcnt\@ne
  \hskip 1\arraycolsep \hfil${##}$\hfil
  &\global\@eqcnt\tw@ \hskip 1\arraycolsep
$\displaystyle\tabskip\z@{##}$
\hfil  \tabskip\@centering&\global\@eqcnt\thr@@\llap{##}\tabskip\z@
\cr}

\def\endeqnarray{\@@eqncr\egroup
      \global\advance\c@equation\m@ne$$\global\@ignoretrue}

\def\@eqnnum{\hbox to 3cm{\phantom{\rm(\theequation)} \draftnote
                         \hfil}\hskip -3cm {\rm(\theequation)}}

\def\@@eqncr{\let\@tempa\relax
    \ifcase\@eqcnt \def\@tempa{& & &}\or \def\@tempa{& &}
      \or \def\@tempa{&}
      \or\def\@tempa{}
\fi\@tempa
\if@eqnsw
\if@fewtab\@eqnnum\fi
\stepcounter{equation}\fi\global
\@eqnswtrue\global\@eqcnt\z@\global\@fewtabtrue\cr}

\def\draftcite#1{\ifnum\draftcontrol=1#1\else{}\fi}

\def\@lbibitem[#1]#2{\item{}\hskip -3cm \hbox to 2cm
{\hfil$\scriptstyle\draftcite{#2}$}\hskip
1cm[\@biblabel{#1}]\if@filesw
     {\def\protect##1{\string ##1\space}\immediate
      \write\@auxout{\string\bibcite{#2}{#1}}}\fi\ignorespaces}

\def\@bibitem#1{\item\hskip -3cm \hbox to 2cm
{\hfil $\scriptstyle\draftcite{#1}$}\hskip 1cm
\if@filesw \immediate\write\@auxout
       {\string\bibcite{#1}{\the\value{\@listctr}}}\fi\ignorespaces}

\def\nsection#1{\section{#1}\setcounter{equation}{0}}

\font\tendl=msbm10  scaled \magstep1
\font\sevendl=msbm7 scaled \magstep1
\font\fivedl=msbm5 scaled \magstep1
\font\tengl=eufm10  scaled \magstep1
\font\sevengl=eufm7 scaled \magstep1
\font\fivegl=eufm5 scaled \magstep1

\newfam\dlfam \def\dl{\fam\dlfam\tendl} 
\textfont\dlfam=\tendl \scriptfont\dlfam=\sevendl
\scriptscriptfont\dlfam=\fivedl
\newfam\glfam  
\textfont\glfam=\tengl \scriptfont\glfam=\sevengl
\scriptscriptfont\glfam=\fivegl

\def\draftdate{\number\month/\number\day/\number\year\ \ \ \hourmin }

\global\def\draftcontrol{0}
\catcode`\@=12
\def\tilde{\widetilde}

\documentstyle[11pt]{article}

\def\theequation{{\thesection.\arabic{equation}}}

\newcommand{\be}{\begin{eqnarray}}
\newcommand{\en}{\end{eqnarray}\vs 0.5 cm}

\newcommand{\no}{\noindent}
\newcommand{\vs}{\vskip}
\newcommand{\hs}{\hspace}

\newcommand{\NR}{{{\dl R}}}
\newcommand{\NP}{{{\dl P}}}
\newcommand{\NC}{{{\dl C}}}
\newcommand{\qq}{\begin{eqnarray}}
\newcommand{\de}{\bar\partial}
\newcommand{\da}{\partial}
\newcommand{\ee}{{\rm e}}

\newcommand{\qqq}{\end{eqnarray}}

\newcommand{\tr}{\hbox{tr}}

\newcommand{\CA}{{\cal A}}

\newcommand{\CC}{{\cal C}}

\newcommand{\CG}{{\cal G}}

\newcommand{\CO}{{\cal O}}

\newcommand{\s}{\hspace{0.05cm}}
\newcommand{\m}{\hspace{0.025cm}}

\newcommand{\bx}{{\bf x}}
\newcommand{\bv}{{\bf v}}
\newcommand{\bu}{{\bf u}}
\newcommand{\bq}{{\bf q}}
\begin{document}
\title{\bf Coulomb gas representation of the SU(2) WZW
correlators at higher genera}
\author{\ \\Krzysztof Gaw\c{e}dzki \\ C.N.R.S.,
I.H.E.S., 91440 Bures-sur-Yvette, France}
\date{ }
\maketitle

\vskip 1 cm

\begin{abstract}
We  extend the analysis of \cite{Gaw1} to the case
with insertion points. The result allows
to express the correlation functions of the
\s$SU(2)\s$ WZW conformal field theory
on Riemann surfaces of genus \s$>\m 1\s$ by finite
dimensional integrals.
\end{abstract}
\vskip 1.7cm

\nsection{\hspace{-.6cm}.\ \ Introduction}
\vs 0.5cm

The present paper completes the work \cite{Gaw1}
where we have computed the scalar product
of the \s$SU(2)\s$ Chern-Simons states on a
Riemann surface of genus \s$>\m 1\s$ in the
absence of Wilson lines. Here, we treat
the case with Wilson lines \s$\CC_l\m$, \s
in representations of spin \s$j_l\s$, \s
cutting the Riemann surface at marked points.
Let us briefly list the basic notations, definitions
and relations, referring to  \cite{Gaw1} for
a more extensive introduction. Below:
\vs 0.4cm
\s$\Sigma\s$ \s denotes the Riemann surface (of genus
\s$\gamma>1\s$) \s with distinct marked points \s$\xi_l,
\ l=1,\dots,L\s$.
\vs 0.4cm
\s$\CA^{01}\s$ \s stands
for the space of smooth \s$sl(2,\NC)$-valued
\s$0,1$-forms \s$A^{01}\s$ on \s$\Sigma\s$.
\vs 0.4cm
\s$\CG^\NC\s$ \s is the group of complex gauge
transformations given by smooth maps \s$h:\CA^{01}
\longrightarrow SL(2,\NC)\s$ which act by
\s$A^{01}\longmapsto{}^h\hs{-0.1cm}A^{01}\equiv
hA^{01}h^{-1}+h\de h^{-1}\s$ on \s$\CA^{01}\s$.
\vs 0.4cm
\s$S(h,A^{10}+A^{01})\s$ \s denotes  the WZW  model action
in the presence of the gauge field.
\vs 0.4cm
\s$k=1,2,\dots\s$ \s is the level of the  theory.
\vs 0.4cm
\s$V_j\s$ \s stands for the space of spin \s$j\s$
representation with \s$g\in SL(2,\NC)\s$ acting
on it by linear automorphism \s$g_j\s$.
\vs 0.4cm
\s$\Psi:\CA^{01}\longrightarrow
\otimes_lV_{j_l}\s$
\s is a CS state if it is holomorphic and if
\s${}^h\hs{-0.02cm}\Psi\s=\s\Psi\s$
where
\qq
({}^h\hs{-0.02cm}\Psi)(A^{01})\s=\s\m
\ee^{-k\m S(h,A^{01})}\ \otimes_l h(\xi_l)_{j_l}\s
\Psi({}^{h^{-1}}\hs{-0.15cm}A^{01})\s.
\qqq
\vs 0.4cm
\s$W_k((\xi_l),(j_l))\s$ \s denotes the
(finite-dimensional) space of CS states.
\vs 0.4cm
The scalar product of the CS states is formally
given by the integral
\qq
\|\Psi\|^2\s=\s\int|\Psi(A^{01})|^2_{\otimes V_{j_l}}
\s\ee^{{ik\over2\pi}\smallint_\Sigma
\tr\s\s\m A^{10}\wedge A^{01}}
\s DA\ .
\label{FIn}
\qqq
The aim of the paper is to compute the functional
integral (\ref{FIn}) over the \s$su(2)\s$ gauge fields
\s$A=A^{10}+A^{01}\s$ with \s$A^{10}=-(A^{01})^\dagger\s$
by reducing it to an explicit finite-dimensional
integral. Such a reduction allows
to express the correlation functions
in the external gauge field of
the WZW model of conformal field theory,
\qq
\Gamma((\xi_l),(j_l),A)\s=\s\int\otimes_l g(\xi_l)_{j_l}
\s\ee^{-k\m S(g,A)}\s\s Dg\ ,
\qqq
by finite-dimensional integrals. According to \cite{Quadr},
\qq
\Gamma((\xi_l),(j_l),A)\s=\s\sum\limits_{r,r'}
H^{rr'}\s\Psi_r(A^{01})\otimes
{\overline{\Psi_{r'}(A^{01})}}\ \m\ee^{{ik\over 2\pi}
\smallint\tr\s\s A^{10}\wedge A^{01}}
\qqq
for any basis \s$(\Psi_r)\s$ of \s$W_k((\xi_l),(j_l))\s$
with matrix \s$(H^{rr'})\s$ inverting the matrix
\s$((\Psi_r,\Psi_{r'}))\s$ of scalar products of \s$\Psi_{r}$'s.
Hence
\qq
&\displaystyle{\Gamma((\xi_l),(j_l),A)\s=\s\sum\limits_{r,r'}
\s\m\Psi_r(A^{01})\otimes
{\overline{\Psi_{r'}(A^{01})}}\ \m\ee^{{ik\over 2\pi}
\smallint\tr\s\s A^{10}\wedge A^{01}}\hs{2cm}\ }&\cr
&\cdot\s\int z^r\bar z^{r'}
\m\exp[{-\sum\limits_{{s,s'}}\bar z^s
\m(\Psi_s,\Psi_{s'})\m z^{s'}}]
\prod\limits_td^2z_t\ \bigg/\
\int\exp[{-\sum\limits_{{s,s'}}\bar z^s\m(\Psi_s,\Psi_{s'})
\m z^{s'}}]\prod\limits_td^2z_t&
\qqq
into which one should substitute
the finite-dimensional integral expressions
for the scalar products \s$(\Psi_s,\Psi_{s'})\s$.
\vs 0.5cm

The functional integral in Eq.\s\s(\ref{FIn}) is calculated
by a change of variables \s$A^{01}\s\mapsto\s(h,n)\s$,
\qq
A^{01}=\m{}^{h^{-1}}\hs{-0.13cm}A^{01}(n)\s,
\qqq
where \s$h\in\CG^\NC\s$ and \s$n\s$ parametrizes a
slice \s$\{A^{01}(n)\}\s$
(of complex dimension \s$3(\gamma-1)\equiv N\s$) \s in
\s$\CA^{01}\s$ which cuts a generic
orbit of \s$\CG^\NC\s$ a finite number, say \m$\nu\m$,
\m of times.
Upon the change of variables, Eq.\s\s(\ref{FIn})
becomes
\qq
\|\Psi\|^2\s=\s {_1\over^\nu}\int|\otimes_lh(\xi_l)^{-1}_{j_l}
\m\Psi(A^{01}(n))\m|^2_{\otimes V_{j_l}}
\ \s\ee^{-{{ik}\over{2\pi}}
\smallint_{_\Sigma}
{\rm tr}\s\s (A^{01}(n))^\dagger\wedge A^{01}(n)}
\ \s\ee^{\m(k+4)\m S(hh^\dagger,\s A(n))}
\ \cr
\cdot\ \s{\rm det}\s(\bar D_n^\dagger\s\bar D_n\m)
\ \s{\rm det}\m(\Omega(1,n))^{-1}\s\s
\m|\s{\rm det}\m(\m
\smallint_{_\Sigma}\m\tr\m\s\s\omega^\beta(n)
\wedge{_{\da A^{01}(n)}
\over^{\da n_\alpha}}\m)\m|^2
\s\s\m D(hh^\dagger)\prod\limits_\alpha
d^2n_\alpha\ ,\ \ \ \
\label{newI}
\qqq
where \s$|\s\m\cdot\m\s|_{\otimes V_{j_l}}\s$ is the
norm induced from the scalar product of spaces \s$V_{j_l}\s$,
\s$\bar D_n\equiv\de+[A^{01}(n)\m,\s\cdot\s\m]\s$
acts on \s$sl(2,\NC)$-valued functions and
the matrix \s$\Omega(1,n)=(\s i\smallint_\Sigma\tr\s\m\omega^\alpha(n)
\wedge\omega^\beta(n)^\dagger\s)\s$ with \s$\omega^\alpha(n)\s$,
\s$\alpha=1,\dots,N\s$, \s
running through a basis of \s$sl(2,\NC)$-valued \s$1,0$-forms
annihilated by the dual of \s$\bar D_n\s$.
The passage from Eq.\s\s(\ref{FIn}) to (\ref{newI})
has been discussed in details in the beginning
of Sec.\s\s6 of \cite{Gaw1}.
\vs 0.5cm

We shall use the same slice \s$\{A^{01}(n)\}\s$
as in \cite{Gaw1}. Let us briefly recall its
construction. It is based on representing
the gauge fields \s$A^{01}\s$ in the trivial
bundle \s$\Sigma\times\NC^2\s$ by gauge fields
\s$B^{01}\s$ in the bundle \s$L_0^{-1}\oplus L_0\s$
where \s$L_0\s$ is a fixed holomorphic line bundle
over \s$\Sigma\s$ of degree \s$g-1\s$ s.\s t.
\s$L_0^2\not=K\s$, \s where \s$K\s$ is the canonical
bundle (of \s$1,0$-covectors) on \s$\Sigma\s$.
We shall provide \s$L_0\s$ with a hermitian
structure. \s$\Sigma\times\NC^2\s$ and
\s$L_0^{-1}\oplus L_0\s$ are equivalent as smooth
rank two bundles with trivial determinant
and hermitian structures. Choosing the equivalence
\s\s$U:L_0^{-1}\oplus L_0\longrightarrow\Sigma\times
\NC^2\s$, \s we shall put
\qq
B^{01}=U^{-1}\hs{-0.03cm}A^{01}U+U^{-1}\de U\ .
\qqq
$B^{01}\s$ is a \s$0,1$-form with values in
\s${\rm End}(L_0^{-1}\oplus L_0)\s$. \s We shall
consider triangular forms
\qq
B_{\bx,b}^{01}=\left(\matrix{{-a_{\bf x}}&
b\cr0&{a_{\bf x}}}\right)\ ,
\qqq
where \s$b\s$ is a \s$0,1$-form with the values in
\s$L_0^{-2}\s$ (i.\s e. \s$b\in\wedge^{01}(L_0^{-2})\s$)
\s and
\qq
a_\bx\s=\s\pi\s(
\smallint_{_{{x}_0}}^{^{{x}}}\omega)\s
({\rm Im}\s\tau)^{-1}\m\bar\omega
\label{atilde}
\qqq
in the shorthand notation using the vector
\s$\omega\equiv(\omega^i)\s$ of holomorphic
\s$1,0$-forms and the period matrix \s$\tau\equiv
(\tau^{ij})\s$ corresponding to a fixed marking
of \s$\Sigma\s$. \s$x_0\s$ is a fixed point
on \s$\Sigma\s$ and \s$\bx\s$ is the point in the covering
space \s$\tilde\Sigma\s$ corresponding to the path
from \s$x_0\s$ to \s$x\s$ used in the integral
in (\ref{atilde}). The slice \s$\{A^{01}(n)\}\s$
of \s$\CA^{01}\s$
is obtained by choosing (in a piece-wise holomorphic way)
the gauge fields \s$A^{01}_{\bx,b}\s$ corresponding
to \s$B^{01}_{\bx,b}\s$ with one \s$\bx\s$ for
each \s$x\in\Sigma\s$ and one \s$b\s$ in each
complex ray in \s$\wedge^{01}(L_0^{-2})/
(\de-2a_\bx)(\CC^\infty(L_0^{-2}))\s$. \s The latter
space may be identified with the cohomology
space \s$H^1(L_\bx^{-2})\s$ where \s$L_\bx\s$
denotes the line bundle \s$L_0\s$ with
the holomorphic structure given by the
operator \s$\de_{_{\hs{-0.07cm}L_{\bf x}}}\equiv\de+a_\bx\s$.
\s${\rm dim}(H^1(L^{-2}_\bx))=
N\s(\m\equiv3(\gamma-1)\m)\s$, \s
the number equal to the dimension of the slice
(the one dimension subtracted by considering
projectivized \s$H^1(L^{-2}_\bx)\s$
is added by changing \s$x\in\Sigma\m$).
\vskip 1.7cm

\nsection{\hspace{-.6cm}.\ \ CS states with insertions}
\vs 0.5cm

We shall represent the CS states \s$\Psi\s$
by functions \s$\psi\s$ of \s$\bx\in\tilde\Sigma\s$
and of \s$b\in\wedge^{01}(L_0^{-2})\s$, with values
in \s$\otimes_l V_{j_l}\s$.
To this end,
it will be convenient to realize each
\s$V_{j_l}\s$ as the space of polynomials \s$P(u_l)\s$
of degree \s$\m\leq 2j_l\s$ of variable
\s$u_l\s$ taking values in the fiber
\s$(L_0^{-2})_{\xi_l}\s$ of \s$L_0^{-2}\s$.
\s It will be done so
that the action of
the element \s$U(\matrix{_a&_b\cr^c&^d})^{^{-1}}U^{-1}\s$
of \s$SL(2,\NC)\s$, \s where \s$(\matrix{_a&_b\cr^c&^d})\s$
is a linear endomorphism with determinant one
of \s$(L_0^{-1}\oplus L_0)_{\xi_l}\s$ is given by
\qq
P(u_l)\ \longmapsto\ (cu_l+d)^{2j}\s P((au_l+b)/(cu_l+d))\ .
\qqq
The scalar product in \s$V_{j_l}\s$ is then
given by
\qq
|P|^2\s=\s{_{2j_l+1}\over^\pi}
\int\limits_\NC|P(u_l)|^2\s(1+|u_l|^2)^{-2j_l-2}\s
d^2u_l\ ,\label{spr}
\qqq
where \s$|u_l|^2\s$ is the hermitian square of \s$u_l\s$.
In such a polynomial realization of
\s$\otimes_l V_{j_l}\s$,
\s$\psi(\bx,b)\s$ will be defined by the relation:
\qq
\psi(\bx,b)(\bu)\s=\s\m\exp[\s{_{ik}\over^{2\pi}}
\smallint_{_\Sigma}\tr\s\s A^{10}_0\hs{-0.1cm}\wedge
\hs{-0.08cm}A^{01}_{{\bf x},b}]\s\ \s
\Psi(A^{01}_{{\bf x},b})(\bu-\bv(b))\ ,
\label{newCS}
\qqq
where \s$\bu\equiv(u_l)\s$ and \s$\bv(b)\equiv(v(b)(\xi_l))\s$
with
\qq
v(b)(\xi)\s\equiv\s(\de_{_{\hs{-0.07cm}L_{\bf x}^{-2}}}^{\s\s-1}
\m b)(\xi)\ .
\label{ub}
\qqq
In the latter formula, \s$\de_{_{\hs{-0.07cm}L_{\bf x}^{-2}}}^{\s\s-1}\s$
stands for the orthogonal projection (in the L$^{\rm 2}$
scalar product) on the image of the operator
\s\s$\de_{_{\hs{-0.07cm}L_{\bf x}^{-2}}}\equiv
\de-2a_\bx:\CC^\infty(L_0^{-2})
\m\longrightarrow\m\wedge^{01}(L_0^{-2})\s\s$
followed by the inverse of \s$\de_{_{\hs{-0.07cm}L_{\bf x}^{-2}}}\s$
mapping onto its image. Note that, for
\s$v\in\CC^\infty(L_0^{-2})\s$,
\qq
v(b+\de_{_{\hs{-0.07cm}L_{\bf x}^{-2}}}\m v)=v(b)+v\ .
\qqq
Generalizing the arguments of Sect.\s\s 3 of \cite{Gaw1}
to the present case, one obtains, as a consequence
of the gauge invariance of the CS states, the following
relations
\qq
&&\hbox to 3.2cm{$\psi(\bx,b+\de_{_{\hs{-0.07cm}L_{\bx}^{-2}}}\m v)
(\bu)$\hfill}\s=\s\s\psi(\bx,b)
(\bu)\ ,\cr
&&\hbox to 3.2cm{$\psi(\bx,\lambda b)(\lambda\bu)$\hfill}
\s=\s\s\lambda^{k(\gamma-1)+J}
\s\psi(\bx,b)(\bu)\ ,\cr
&&\hbox to 3.2cm{$\psi(p\bx,c_p^2\m b)(\bu)$\hfill}
\s=\s\s\mu(p,\bx)^k\s\nu(c_p)^k
\s\prod\limits_lc_p(\xi_l)^{2j_l}
\s\s\psi(\bx,b)(\m c_p(\xi_l)^{-2}u_l)
\qqq
for \s$v\in\CC^\infty(L_0^{-2})\s$, \s$\lambda\in\NC\s$,
\s$J\equiv\sum_lj_l\s$,
\s$p\in\Pi_1(\Sigma,x_0)\s$ and \s\s$c_p\s$,
\s$\mu(p,\bx)\s$ and \s$\nu(c_p)\s$ as in
Sect.\s\s 3 of \cite{Gaw1}. In particular, writing
\qq
\psi(\bx,b)(\bu)\s=\s\sum\limits_{\bq\equiv(q_l)}
\psi_{\bq}(\bx,b)\s\prod\limits_lu_l^{q_l}
\qqq
with \s$q_l=0,1,\cdots,2j_l\s$, we infer that
\s$\psi_\bq(\bx,\s\cdot\s)\s$ is a homogeneous
polynomial on \s$H^1(L_\bx^{-2})\s$ of degree
\s$k(\gamma-1)+J-Q\s$, \s where \s$Q\equiv
\sum_lq_l\s$, \s with values in \s$\otimes_l(L_0^{2q_l}
)_{\xi_l}\s$. \s Moreover,
\qq
\psi_{\bf q}
(p\bx,c_p^2\m b)(\bu)\s=\s\mu(p,\bx)^k\m\nu(c_p)^k
\m\prod\limits_lc_p(\xi_l)^{2(j_l-q_l)}
\m\s\psi_{\bf q}(\bx,b)\ .
\qqq
For the interpretation of the last relation, one should
remark that the function \s$\Phi\s$ on \s$\tilde\Sigma\s$
transforming under the action of \s$\Pi_1(\Sigma,x_0)\s$
as
\qq
\Phi(p\bx)\s=\s\mu(p,\bx)^k\m\nu(c_p)^k
\m\prod\limits_lc_p(\xi_l)^{2(j_l-q_l)}\s\m\Phi(p\bx)
\qqq
defines a holomorphic section of the bundle
\s$L_0^{2k}K^k(2\sum_l(j_l-q_l)\xi_l+2(k(2-\gamma)-J+Q)x_0)\s$
($\m MN(D)\equiv M\otimes N\otimes\CO(D)\s$ for line
bundles \s$M,\ N\s$ and a divisor \s$D$\s).
In the notation of Sect.\s\s3 and 4 of \cite{Gaw1},
\s$\psi_\bq\s$ is a holomorphic section of the bundle
\qq
&\varpi^*(L^{2k}K^k(2\sum\limits_l(j_l-q_l)\xi_l+2(k(1-\gamma)
-J+Q)x_0))\
{\rm Hf}(W_0)^{k(1-\gamma)-J+Q}&\cr
&\s\cong\ \s
\varpi^*(L^{2k}K^k(2\sum\limits_l(j_l-q_l)\xi_l))\ {\rm Hf}
(W_0')^{k(1-\gamma)-J+Q}&
\nonumber
\qqq
with values in \s$\otimes_l(L_0^{2q_l})_{\xi_l}\s$.
\vskip 1.7cm

\nsection{\hspace{-.6cm}.\ \ Scalar product formula}
\vs 0.5cm

With the use of the slice of \s$\CA^{01}\s$ described
above, the functional integral (\ref{newI})
reduces to the integral over fields \s$U^{-1}h\m U\s
=\s(\matrix{_{\ee^{\varphi/2}}&_{\ee^{\varphi/2}\m w}\cr
^0&^{\ee^{-\varphi/2}}})\s$ (\s$\varphi\s$
is a real function on \s$\Sigma\s$, \s$w\s$ is a
section of \s$L_0^{-2}\s$)\m, \s over the projective
space \s$\NP H^1(L_\bx^{-2})\s$ and over \s$\bx\s$
belonging to a fundumental domain in \s$\tilde\Sigma\s$.
The functional integration will be performed essentially
as in \cite{Gaw1}. In the first step, one rewrites
the expression under the integral in Eq.\s\s(\ref{newI})
in an explicit way (Sect. 6.1 to 6.3 and Sect.\s\s7
of \cite{Gaw1}\m). The only place where the arguments
of \cite{Gaw1} have to be modified is the treatment of
the term
\qq
|\otimes_lh(\xi_l)^{-1}_{j_l}\s
\Psi(A^{01}(n))\m|^2_{\otimes V_{j_l}}
\qqq
substituting for \s$|\Psi(A^{01}(n)\m|^2\s$ treated
in Sect.\s\s6.1 therein. The modification replaces
\s$|\psi(\bx,b)|^2\s$ at the end of Eq.\s\s(6.17)
of \cite{Gaw1} by
\qq
\prod\limits_l{_{2j_l+1}\over^\pi}\int
|\s\psi(\bx,b)\m(\m\ee^{\varphi(\xi_l)}
(u_l+w(\xi_l))\m+\s v(b)(\xi_l)\m)\s|^2\
\prod\limits_l{_{\ee^{-2j_l\varphi(\xi_l)}\s d^2u_l}
\over^{(1+|u_l|^2)^{2j_l+2}}}\ ,
\qqq
where \s$v(b)\s$ is given by Eq.\s\s(\ref{ub}).
Altogether, the explicit form of Eq.\s\s(\ref{newI})
is
\qq
\|\Psi\|^2\ =\s\ {\rm const}.\s\left({_{{\rm det}'(-\Delta)}
\over^{{\rm area}\m\cdot\m{\rm det}({\rm Im}\m\tau)}}\right)
\ \ee^{-{ik\over2\pi}\smallint_\Sigma{\rm tr}
\s\s A^{10}_0\wedge A^{01}_0}\s\int\s
\ee^{-2\pi k\m(\smallint_{x_0}^{\bf x}\bar\omega)\m
{1\over{\rm Im}\m\tau}\m
(\smallint_{x_0}^{\bf x}\omega)}\hs{1.5cm}\ \ \ \cr\cr
\cdot\ {\rm det}(\smallint_{_\Sigma}\langle\eta^\alpha_\bx,
\m\wedge\eta^\beta_\bx\rangle)^{-1}\ {\rm det}
(\smallint_{_\Sigma}\langle\kappa^r_\bx,\m\kappa^s_\bx
\rangle\s{\rm vol}\m)^{-1}\ {\rm det}
(\de_{_{\hs{-0.07cm}L_{\bf x}^{-2}}
}^{\s\s\dagger}\m\de_{_{\hs{-0.07cm}L_{\bf x}^{-2}}}\m)\
{\rm det}'(\de_{_{\hs{-0.07cm}L_{\bf x}^2}}^{\s\s\dagger}
\m\de_{_{\hs{-0.07cm}L_{\bf x}^2}}\m)\hs{1cm}\ \ \ \cr\cr
\cdot\ |\sum\limits_{j=1}^g(-1)^j\s\m{\rm det}
(\smallint_\Sigma\kappa_\bx^r\m\omega^i\wedge
b)_{_{i\not=j}}\s\omega^j(x)\s|^{^{\wedge 2}}\s\m
\bigg(\smallint_{\NC^N}|\s\psi(\bx,b)\m(\m\ee^{\varphi(\xi_l)}
(u_l+w(\xi_l))+v(b)(\xi_l)\m)\s|^2\ \ \ \cr\cr
\cdot\s\prod\limits_l{_{
\ee^{-2j_l\varphi(\xi_l)}\s d^2u_l}
\over^{(1+|u_l|^2)^{2j_l+2}}}\bigg)\s
\s\s\ee^{{i(k+4)\over 2\pi}
\int_{_\Sigma}[\s\varphi(\da\de\varphi
-2F_0)\s-\s\langle\ee^{-\varphi}b+
(\de_{_{\hs{-0.07cm}L_{\bf x}^{-2}}}
+\de\varphi)w\m,
\m\wedge(\ee^{-\varphi}b+(\de_{_{\hs{-0.07cm}L_{\bf x}^{-2}}}
+\de\varphi)w)\rangle\s]}\ \ \ \ \cr
\cdot\ |\epsilon_{\alpha_1,\m\dots\m,\m\alpha_N}\s
z_\bx^{\alpha_1}\m dz_\bx^{\alpha_2}\wedge\dots\wedge
dz_\bx^{\alpha_n}
\m|^{^{\wedge 2}}\ Dw\ D\varphi\s\m,\ \ \hs{0.5cm}
\label{expl}
\qqq
where \s$(\kappa^r_\bx)_{_{r=1}}^{^{\gamma-1}})\s$
is a basis of \s$H^0(L_\bx^2)\s$, \s$(
\eta^\alpha_\bx)_{_{
\alpha=1}}^{^{N}}\s$ is a basis of \s$H^0(L_\bx^2
K)\s$, \s$z^\alpha_\bx=\smallint_{_\Sigma}
\eta^\alpha_\bx\wedge b\s$ provide homogeneous coordinates
on \s$\NP H^1(L_\bx^{-2})\s$, \s$\langle\s\cdot\s,\s\cdot\s
\rangle\s$ denotes the hermitian structures on powers
of the bundle \s$L_\bx\s$ induced by the fixed hermitian
metric on \s$L_0\s$ (of curvature \s$F_0\s$)
and \s${\rm vol}\s$ is the Riemannian
volume on \s$\Sigma\s$. \s We shall
shift above the field \s$w\s$ by \s$-\ee^{-\varphi}v(b)\s$,
see Eq.\s\s(\ref{ub}). Note that
\qq
b-\de_{_{\hs{-0.07cm}L_{\bf x}^{-2}}}v(b)\s=\s i\s z^\alpha_\bx\m(H_0^{-1}
)_{_{\alpha\beta}}\m{\eta^{\beta}_\bx}^{\dagger}
\label{ostat}
\qqq
where \s$(H_0)^{^{\alpha\beta}}\s=\s{_1\over^i}
\smallint_{_{\Sigma}}\langle\eta^\alpha_\bx,\m\wedge
\eta^\beta_\bx\rangle\s$ (the right hand side of
Eq.\s\s(\ref{ostat}) gives the component of \s$b\s$
orthogonal to \s$\de_{_{\hs{-0.07cm}L_{\bf x}^{-2}}}(\CC^\infty
(L_\bx^{-2}))\s$)\m. \s The shift of \s$w\s$
removes \s$\bv(b)\s$ from the argument of \s$\psi(\bx,b)\s$
and replaces \s$\ee^{-\varphi}b+(\de_{_{\hs{-0.07cm}L_{\bf x}^{-2}}}
+\de\varphi)w\s$ by
\qq
d\s\equiv\s i\s\ee^{-\varphi}\m z^\alpha_\bx\m(H_0^{-1}
)_{_{\alpha\beta}}\m{\eta^{\beta}_\bx}^{\dagger}\s+\s
(\de_{_{\hs{-0.07cm}L_{\bf x}^{-2}}}+\de\varphi)w\ .
\qqq
It will be convenient to decompose
\qq
\varphi=a+\tilde\varphi\ ,\hs{1cm}
D\varphi\s=\s{\rm area}^{^{1/2}}\s\m da\ D\tilde\varphi\ ,
\qqq
where \s$a\in\NR\s$ is the constant contribution
to \s$\varphi\s$ and \s$\tilde\varphi\s$ is orthogonal
to the constant mode. We shall also multiply
the right hand side of Eq.\s\s(\ref{expl}) by
\s$1=\smallint_0^{2\pi}d\theta/2\pi\s$ and
replace the variables \s$z^\alpha_\bx\s$ by \s$\ee^{-i\theta}
z^\alpha_\bx\s$. Finally, we shall perform the change
of variables
\qq
(\theta,\m a,\m z^{\alpha_2}_\bx,\m\dots\m,\m
z^{\alpha_N}_\bx,\m w)\ \longmapsto\ (\zeta^{1}_\bx\equiv
\ee^{-a-i\theta}z^1_\bx,\s\m\dots\m,\s\zeta^N_\bx\equiv
\ee^{-a-i\theta}z^N_\bx\m,\s\m w)\ \cr
\ \longmapsto\ d\s=
\s i\s\ee^{-\tilde\varphi}\m\zeta^\alpha_\bx\m(H_0^{-1}
)_{_{\alpha\beta}}\m{\eta^{\beta}_\bx}^{\dagger}\m+\m
(\de_{_{\hs{-0.07cm}L_{\bf x}^{-2}}}+\de\tilde\varphi)w\ \in\s\wedge^{01}
(L_\bx^{-2})\ .
\label{ChVar}
\qqq
The corresponding
Jacobian is easily calculated to be
\qq
&\ee^{2aN}\s|z^{\alpha_1}_\bx|^{-2}
\s\s{\rm det}(H_{\tilde\varphi})
\ \s{\rm det}\bigg((\de_{_{\hs{-0.07cm}L_{\bf x}^{-2}}}
+\de\tilde\varphi)^\dagger
\s(\de_{_{\hs{-0.07cm}L_{\bf x}^{-2}}}+\de\tilde\varphi)
\bigg)^{\hs{-0.1cm}
^{-1}}&\cr
&=\s\ee^{2aN}\s|z^{\alpha_1}_\bx|^{-2}\s\s
{\rm det}(H_{0})\ \s{\rm det}
\m(\m\de_{_{\hs{-0.07cm}L_{\bf x}^{-2}}}^{\s\s\dagger}
\s\de_{_{\hs{-0.07cm}L_{\bf x}^{-2}}}\m)^{-1}\
\s\ee^{\m{1\over \pi i}\smallint_{_\Sigma}
\tilde\varphi\m(\da\de\tilde\varphi
-2F_0)\s+\s{1\over 2\pi i}
\smallint_{_\Sigma}\tilde\varphi\m R}&
\qqq
where \s\s$(H_{\tilde\varphi})^{^{\alpha\beta}}\s=\s
{_1\over^i}\smallint_{_{\Sigma}}\ee^{\m 2\tilde\varphi}
\langle\eta^\alpha_\bx,\m\wedge
\eta^\beta_\bx\rangle\s\s$
and \s$R\s$ denotes the metric
curvature of \s$\Sigma\s$. \s
In the last line we have used the chiral anomaly
to extract the explicit dependence of the determinants
on \s$\tilde\varphi\s$. \s
Implementing all the above transformations
in Eq.\s\s(\ref{expl}), we obtain
\qq
\|\Psi\|^2\s\m=\s\ {\rm const}.\left({_{{\rm det}'(-\Delta)}
\over^{{\rm area}^{1/2}\m\cdot\m{\rm det}
({\rm Im}\m\tau)}}\right)
\s\ee^{-{ik\over2\pi}\smallint_\Sigma{\rm tr}
\s\s A^{10}_0\wedge A^{01}_0}\s \int\m
\ee^{-2\pi k
\m(\smallint_{x_0}^{\bf x}\bar\omega)\m
{1\over{\rm Im}\m\tau}\m
(\smallint_{x_0}^{\bf x}\omega)}\ \ \ \ \cr\cr
\cdot\ \s{\rm det}(\smallint_{_\Sigma}\langle
\kappa^r_\bx,\m\kappa^s_\bx
\rangle\s{\rm vol}\m)^{-1}\
\s{\rm det}'(\de_{_{\hs{-0.07cm}L_{\bf x}^2}}^{\s\s\dagger}
\m\de_{_{\hs{-0.07cm}L_{\bf x}^2}}\m)\ \s
|\sum\limits_{j=1}^g(-1)^j\s\m{\rm det}(\s\zeta^\alpha_\bx
\m(H_0^{-1})_{_{\alpha\beta}}
\s\s M^{ri\beta})_{_{i\not=j}}\ \ \ \ \cr
\cdot\s\m\omega^j(x)\s|^{^{\wedge 2}}
\s\bigg(\smallint_{\NC^N}|\s\psi(\bx\m,\s
\zeta_\bx)\m(\m\ee^{\tilde\varphi(\xi_l)}
(u_l+w(\xi_l))\m)\s|^2\ \prod\limits_l{_{
\ee^{-2j_l\tilde\varphi(\xi_l)}\s d^2u_l}
\over^{(1+|u_l|^2)^{2j_l+2}}}\bigg)\ \ \ \ \cr\cr
\cdot\ \s\ee^{{i(k+2)\over 2\pi}
\smallint_\Sigma\s\tilde\varphi\m(\da\de\tilde\varphi
-2F_0)\s+{1\over 2\pi i}\smallint_{_\Sigma}\tilde\varphi
\m R\s-\s{i(k+4)\over 2\pi}\m\langle d\m,\s d\rangle\s}\
\s Dd\ D\tilde\varphi\s\m,\ \ \ \s
\label{expl1}
\qqq
where we have denoted:
\qq
M^{ri\beta}\s\equiv\s\smallint_{_\Sigma}\kappa^r_\bx\m\omega^i
\wedge{\eta^\beta_\bx}^\dagger\ ,\hs{0.8cm}
\psi(\bx\m,\s\zeta_\bx)\s\equiv\s\psi(\bx\m,\s
i\m\zeta^\alpha_\bx(H_0^{-1})_{_{\alpha\beta}}
{\eta^\beta_\bx}^\dagger)\ .
\qqq
Observe that the zero mode \s$a\s$ of \s$\varphi\s$ has
been completely absorbed into \s$\zeta_\bx\s$. \s
The change of variables (\ref{ChVar}) is easy to invert.
The relation
\qq
\ee^{\tilde\varphi}d\s=\s i\m\zeta^\alpha_\bx\m(H_0^{-1}
)_{_{\alpha\beta}}\m{\eta^\beta_\bx}^\dagger\s+\s
\de_{_{\hs{-0.07cm}L_{\bf x}^{-2}}}\m v
\qqq
where \s$v\equiv\ee^{\tilde\varphi}w\s$ implies that
\qq
v\s=\s\de_{_{\hs{-0.07cm}L_{\bf x}^{-2}}}^{\s\s{-1}}
\s\ee^{\tilde\varphi}d\hs{0.5cm}{\rm and}
\hs{0.5cm}\zeta^\alpha_\bx\s=\s\smallint_{_\Sigma}
\eta^\alpha_\bx\wedge\ee^{\tilde\varphi}d\ .\label{dd}
\qqq
The \s$d\s$ integral is now straightforward. Since,
by (\ref{dd}),
\qq
{_{\delta}\over^{\delta d(y)}}\s=\s\bigg({_{\da}\over
^{\da\zeta^\alpha_{\bf x}}}\m\eta^\alpha_\bx(y)
\s+\s\smallint_{_\Sigma}{_{\delta}\over^{\delta v(\xi)}}\m
\de_{_{\hs{-0.07cm}L_{\bf x}^{-2}}}^{\s\s-1}(\xi,y)\bigg)
\m\ee^{\tilde\varphi(y)}\ ,
\qqq
performing the \s$d\s$ integration in (\ref{expl1}) gives:
\qq
&\displaystyle{\|\Psi\|^2\s\m
=\s\ {\rm const}.\left({_{{\rm det}'(-\Delta)}
\over^{{\rm area}^{1/2}\m\cdot\m{\rm det}
({\rm Im}\m\tau)}}\right)
\s\ee^{-{ik\over2\pi}\smallint_\Sigma{\rm tr}
\s\s A^{10}_0\wedge A^{01}_0}\s\s\int\m
\ee^{-2\pi k\m(\smallint_{x_0}^{\bf x}\bar\omega)\m
{1\over{\rm Im}\m\tau}\m
(\smallint_{x_0}^{\bf x}\omega)}\hs{1.3cm}\ \ \ \ }&\cr\cr
&\displaystyle{\cdot\ {\rm det}(\smallint_{_\Sigma}\langle
\kappa^r_\bx,\m\kappa^s_\bx
\rangle\s{\rm vol}\m)^{-1}\
\s{\rm det}'(\de_{_{\hs{-0.07cm}L_{\bf x}^2}}^{\s\s\dagger}
\m\de_{_{\hs{-0.07cm}L_{\bf x}^2}}\m)\ \s
\exp\bigg[\smallint_{_\Sigma}
\m(\s{_{\da}\over
^{\da\zeta^\alpha_{\bf x}}}\m\eta^\alpha_\bx(y)\hs{2.5cm}
\ \ \ \ }&\cr\cr
&\displaystyle{\hs{0.4cm}+\s\smallint_{_\Sigma}
{_{\delta}\over^{\delta v(\xi)}}\s
\de_{_{\hs{-0.07cm}L_{\bf x}^{-2}}}^{\s\s-1}
(\xi,y)\s)\s\m\ee^{2\tilde\varphi(y)}
\s(\s{\eta^\beta_\bx}^\dagger(y)\m{_{\da}\over
^{\da\bar\zeta_\bx^\beta}}\s
+\s\smallint_{_\Sigma}
\de_{_{\hs{-0.07cm}L_{\bf x}^{-2}}}^{\s\s{-1}\m\dagger}
(y,\xi')\m{_{\delta}\over^{\delta\bar v(\xi')}}\s)\bigg]
\bigg|_{{{\zeta_\bx=0}\atop{v=0}}}\ \ \ \ }&\cr\cr
&\displaystyle{\cdot\
|\sum\limits_{j=1}^g(-1)^j\s\m{\rm det}
(\s\zeta^\alpha_\bx\m
\m(H_0^{-1})_{_{\alpha\beta}}\m M^{ri\beta})_{_{i\not=j}}
\s\omega^j(x)\s|^{^{\wedge 2}}
\s\bigg(\smallint_{\NC^N}|\s\psi(\bx\m,\s
\zeta_\bx)\m(\m\ee^{\tilde\varphi(\xi_l)}
u_l\m+\m v(\xi_l)\m)\s|^2\ \ \ \ }&\cr\cr
&\displaystyle{\hs{2cm}\cdot\s\prod\limits_l{_{
\ee^{-2j_l\tilde\varphi(\xi_l)}\s d^2u_l}
\over^{(1+|u_l|^2)^{2j_l+2}}}\bigg)
\s\s\ee^{{i(k+2)\over 2\pi}
\smallint_\Sigma\s\tilde\varphi\m(\da\de\tilde\varphi
-2F_0)\s+{1\over 2\pi i}\smallint_{_\Sigma}\tilde\varphi
\m R}\ \s D\tilde\varphi\s\m.\ \ \ \s}&
\label{expl2}
\qqq
The \s$u_l$-integral may be written
in the following form:
\qq
\smallint_{\NC^N}|\s\psi(\bx\m,\s
\zeta_\bx)\m(\m\ee^{\tilde\varphi(\xi_l)}
u_l\m+\m v(\xi_l)\m)\s|^2\s\prod\limits_l{_{
\ee^{-2j_l\tilde\varphi(\xi_l)}\s d^2u_l}
\over^{(1+|u_l|^2)^{2j_l+2}}}\ \cr
=\s\sum\limits_{\bf p}\s|\m\tilde
\psi_{\bf p}(\bx,\m\zeta_\bx)
(v(\xi_l)\m)\m|^{^2}\s\prod
\limits_l\ee^{-2(j_l-p_l)
\tilde\varphi
(\xi_l)}
\qqq
where \s$\tilde\psi_{\bf p}(\bx,\m\zeta_\bx)(v(\xi_l))\s$
is a homogeneous function of \s$\zeta^\alpha_\bx\s$
and \s$v(\xi_l)\s$ of degree \s$k(\gamma-1)+J-P\s$
with \s$P\equiv\sum_lp_l\s$.
\s Inserting this decomposition into Eq.\s\s(\ref{expl2})
and denoting \s$(k+1)(\gamma-1)\s$ by \s$M\s$,
\s we obtain
\qq
&\displaystyle{\|\Psi\|^2\s\m
=\s\ {\rm const}.\left({_{{\rm det}'(-\Delta)}
\over^{{\rm area}^{1/2}\m\cdot\m{\rm det}
({\rm Im}\m\tau)}}\right)
\s\ee^{-{ik\over2\pi}\smallint_\Sigma{\rm tr}
\s\s A^{10}_0\wedge A^{01}_0}\m\s\int\m
\s\ee^{-2\pi k
\m(\smallint_{x_0}^{\bf x}\bar\omega)\m
{1\over{\rm Im}\m\tau}\m
(\smallint_{x_0}^{\bf x}\omega)}\hs{1cm}\ \ \ \ }&\cr\cr
&\displaystyle{\cdot\ \s{\rm det}(\smallint_{_\Sigma}\langle
\kappa^r_\bx,\m\kappa^s_\bx
\rangle\s{\rm vol}\m)^{-1}\
\s{\rm det}'(\de_{_{\hs{-0.07cm}L_{\bf x}^2}}^{\s\s\dagger}
\m\de_{_{\hs{-0.07cm}L_{\bf x}^2}}\m)\
\s\sum\limits_{\bf p}\bigg[\smallint_{_\Sigma}
\m(\s{_{\da}\over
^{\da\zeta^\alpha_{\bf x}}}\m\eta^\alpha_\bx(y)\hs{2cm}
\ \ \ \ }&\cr\cr
&\displaystyle{+\s\smallint_{_\Sigma}
{_{\delta}\over^{\delta v(\xi)}}\s
\de_{_{\hs{-0.07cm}L_{\bf x}^{-2}}}^{\s\s-1}
(\xi,y)\s)\s\m\ee^{2\tilde\varphi(y)}
\s(\s{\eta^\beta_\bx}^\dagger(y)\m{_{\da}\over
^{\da\bar\zeta_\bx^\beta}}\s+\s\smallint_{_\Sigma}
\de_{_{\hs{-0.07cm}L_{\bf x}^{-2}}}^{\s\s{-1}\m\dagger}
(y,\xi')\m{_{\delta}\over^{\delta\bar v(\xi')}}\s)\bigg]^{{
\hs{-0.07cm}M+J-P}}
\bigg|_{{{\zeta_\bx=0}\atop{v=0}}}\ \ \ \ }&\cr\cr
&\displaystyle{\cdot\
|\sum\limits_{j=1}^g(-1)^j\s\m{\rm det}
(\s\zeta^\alpha_\bx\m
\m(H_0^{-1})_{_{\alpha\beta}}\m M^{ri\beta})_{_{i\not=j}}
\s\omega^j(x)\s|^{^{\wedge 2}}\ \s|\s\tilde
\psi_{\bf p}(\bx\m,\s\zeta_\bx)\m(\m v(\xi_l)\m)\s|^2
\ \ \ \ }&\cr\cr
&\displaystyle{\hs{1.5cm}\cdot\s\prod\limits_{l}
\ee^{-2(j_l-p_l)\tilde\varphi(\xi_l)}\
\s\ee^{{i(k+2)\over 2\pi}
\smallint_\Sigma\s\tilde\varphi\m(\da\de\tilde\varphi
-2F_0)\s+\s{1\over 2\pi i}\smallint_{_\Sigma}\tilde\varphi
\m R}\s\ D\tilde\varphi\s\m.\ \ \ \s}&
\label{expl3}
\qqq
We are still left with the Gaussian integration
over \s$\tilde\varphi\s$. It is of the form
\qq
&\int\exp\m[\sum\limits_{m=1}^{M+J-P}
\hs{-0.1cm}2\tilde\varphi(y_m)\s
-\s\sum\limits_l2(j_l-p_l)
\tilde\varphi(\xi_l)
+{i(k+2)\over 2\pi}
\smallint_\Sigma\s\tilde\varphi\m(\da\de\tilde\varphi
-2F_0)\ &\cr
&+{1\over 2\pi i}\smallint_{_\Sigma}\tilde\varphi
\m R\s]\m\ D\tilde\varphi\s
\equiv\s\int\ee^{-i\smallint_{_\Sigma}\tilde\varphi\m\sigma
+{k+2\over 2\pi i}\int_{_\Sigma}\da\tilde\varphi
\wedge\de\tilde\varphi}\ D\tilde\varphi\ &\cr\cr
&=\s{\rm const}.\ {\rm det}'(-\Delta)^{-1/2}\ \ee^{\m{\pi\over
k+2}\smallint_{_\Sigma}\smallint_{_\Sigma}\sigma(x)\m G(x,y)
\m\sigma(y)}\ ,&
\label{res}
\qqq
where \s$\sigma\s\equiv\s{_{k+2}\over^{\pi}}\m F_0
\s+\s{_1\over^{2\pi}}\m R\s-\s2 i\sum\limits_l
(j_l-p_l)\delta_{\xi_l}\s+\s2i\sum\limits_{m}\delta_{y_m}\s$
and \s$G(x,y)\s$ is a Green function of the Laplacian
\s$\Delta\s$ on \s$\Sigma\s$. Note that
\s$\smallint_{_\Sigma}\sigma=0\s$. For
convenience, we shall choose \s$G(\s\cdot\s,\m\cdot\s)\s$
so that \s$\smallint_{_\Sigma}G(x,\m\cdot\s)\s((k+2)F_0+
R/2)\s=\s0\s$. \s Then the right hand side of
Eq.\s\s(\ref{res}) becomes
\qq
&{\rm const}.\ {\rm det}'(-\Delta)^{-1/2}\
\exp\m[\m-{{4\pi}\over k+2}\m(\sum\limits_{m_1,m_2}G(y_{m_1},
\m y_{m_2})\ &\cr
&-\s\sum\limits_{m,l}(j_l-p_l)\m G(y_m,\m\xi_l)
\s+\s\sum\limits_{l_1,l_2}(j_{l_1}-p_{l_1})\m(j_{l_2}-p_{l_2})
\m G(\xi_{l_1},
\m\xi_{l_2})\s]\ .&
\label{diver}
\qqq
Note the divergent contribution from coinciding
points due to the short distance singularity
\s$G(x,y)\cong{_1\over^{2\pi}}\s\ln\s{\rm dist}(x,y)\s$.
These divergences may be regularized by
Wick ordering, i.e. by splitting the coinciding
points at distance \s$\epsilon\s$, extracting
the most divergent factor \s$\epsilon^{-{2\over k+2}
(M+\sum_lj_l(j_l+1))}\s$
appearing in the term with \s${\bf p}=0\s$
and taking the limit \s$\epsilon\rightarrow0\s$.
The less singular terms with \s${\bf p}\not=0\s$
will be annihilated by this multiplicative
renormalization. As the result, one obtains
the following scalar product formula:
\qq
\|\Psi\|^2\s\m=\s\ {\rm const}.\ {_1\over^{{\rm det}
({\rm Im}\m\tau)}}
\s\left({_{{\rm det}'(-\Delta)}
\over^{{\rm area}}}\right)^{^{\hs{-0.1cm}1/2}}
\s\ee^{-{ik\over2\pi}\smallint_\Sigma{\rm tr}
\s\s A^{10}_0\wedge A^{01}_0}
\s\s\prod\limits_{l_1\not=l_2}
\s\ee^{-{{4\pi}\over k+2}\m j_{l_1}j_{l_2}
\m G(\xi_{l_1},\m\xi_{l_2})}\ \ \ \cr
\cdot\s\prod\limits_l
\ee^{-{{4\pi}\over k+2}\m j_l^2:G(\xi_l,\m \xi_l):}
\s\int\m\ee^{-2\pi k
\m(\smallint_{x_0}^{\bf x}\bar\omega)\m
{1\over{\rm Im}\m\tau}\m
(\smallint_{x_0}^{\bf x}\omega)}
\ \s{\rm det}(\smallint_{_\Sigma}\langle
\kappa^r_\bx,\m\kappa^s_\bx
\rangle\s{\rm vol}\m)^{-1}\
\s{\rm det}'(\de_{_{\hs{-0.07cm}L_{\bf x}^2}}^{\s\s\dagger}
\m\de_{_{\hs{-0.07cm}L_{\bf x}^2}}\m)\ \ \ \ \cr\cr
\cdot\prod\limits_{m=1}^{M+J}\hs{-0.1cm}
\bigg(\smallint_{_\Sigma}
(\m{_{\da}\over
^{\da\zeta^\alpha_{\bf x}}}\m\eta^\alpha_\bx(y_m)
+\smallint_{_\Sigma}{_{\delta}\over^{\delta v(\xi)}}
\de_{_{\hs{-0.07cm}L_{\bf x}^{-2}}}^{\s\s-1}
(\xi,y_m)\s)\wedge
(\s{\eta^\beta_\bx}^\dagger(y_m)\m{_{\da}\over
^{\da\bar\zeta_\bx^\beta}}
+\smallint_{_\Sigma}
\de_{_{\hs{-0.07cm}L_{\bf x}^{-2}}}^{\s\s{-1}\m\dagger}
(y_m,\xi')\m{_{\delta}\over^{\delta\bar v(\xi')}})\bigg)
\ \cr
\cdot\ \bigg(|\sum\limits_{j=1}^g(-1)^j\s\m{\rm det}
(\s\zeta^\alpha_\bx\m
\m(H_0^{-1})_{_{\alpha\beta}}\m
\smallint_{_\Sigma}\kappa^r_\bx\m\omega^i
\wedge{\eta^\beta_\bx}^\dagger)_{_{i\not=j}}
\s\omega^j(x)\s|^{^{\wedge 2}}\ \s|\s\psi(\bx\m,\s
\zeta_\bx)\m(\m v(\xi_l)\m)\s|^2\bigg)\hs{0.3cm}\ \cr\cr
\cdot\s\prod\limits_{m_1\not=m_2}
\s\ee^{-{{4\pi}\over k+2}\m
G(y_{m_1},\m y_{m_2})}\s\s\prod\limits_{m}\s\ee^{
-{{4\pi}\over k+2}\m:G(y_m,\m y_m):}\s\s
\prod\limits_{m,l}\ee^{\m{{4\pi}\over k+2}\m
j_l\m G(y_m,\m\xi_l)}\s\m,\s\hs{0.9cm}
\label{expl4}
\qqq
where \s\s$:G(y,\m y):\s\s\equiv\s
\lim\limits_{y'\rightarrow y}
\s\m(\s G(y,\m y')-{_1\over^{2\pi}}\s
\ln\s{\rm dist}(y,\m y')\s)\s$.
\vs 0.4cm

Eq.\s\s(\ref{expl4})
generalizes the ``Coulomb gas'' representation
(9.5) of \cite{Gaw1}
for the scalar product of CS states to the case with
insertions of the Wilson lines. It reduces to
the latter expression if no lines are
present. The integral in (\ref{expl4}) is over
the modular parameter \s$\bx\s$ running through
a fundamental domain of \s$\tilde\Sigma\s$ and
over the positions \s$y_m\in\Sigma\s$
of \s$(k+1)(\gamma-1)+\sum_lj_l\equiv M+J\s$
screening charges.
As in the case with no insertions, we expect
the integrals to converge only if \s$\psi\s$
describes a (globally defined) CS state. The
resulting scalar product should define
the projectively flat Knizhnik-Zamolodchikov-Bernard
connection \cite{KZ}\cite{Denis}
on the holomorphic bundle whose
fibers are formed by the spaces of CS states.
\vs 1.6cm

\no{\large{\bf Acknowledgements}}
\vs 0.5cm

\no The author would like to
thank The Erwin Schr\"{o}dinger Institute in Vienna
for hospitality and support during the work on this paper.

\vs 2.5cm

\end{document}